\documentclass[aps,10pt,prb,twocolumn,superscriptaddress]{revtex4-2}
\usepackage{graphicx}
\usepackage{amsmath}
\usepackage{amsfonts}
\usepackage{amssymb,mathrsfs}
\usepackage{upgreek}
\usepackage{bbold,yfonts}

\newcommand{\bs}[1]{\boldsymbol{#1}}

\begin{document}

\title{Hydrodynamic approach to electronic transport in graphene: energy relaxation}

\author{B.N. Narozhny}
\affiliation{\mbox{Institut for Theoretical Condensed Matter Physics, Karlsruhe Institute of 
Technology, 76128 Karlsruhe, Germany}}
\affiliation{National Research Nuclear University MEPhI (Moscow Engineering Physics Institute),
  115409 Moscow, Russia}
\author{I.V. Gornyi}
\affiliation{\mbox{Institut for Quantum Materials and Technologies, Karlsruhe Institute of
Technology, 76021 Karlsruhe, Germany}}
\affiliation{Ioffe Institute, 194021 St. Petersburg, Russia}

\date{\today}

\begin{abstract}
  In nearly compensated graphene, disorder-assisted electron-phonon
  scattering or ``supercollisions'' are responsible for both
  quasiparticle recombination and energy relaxation. Within the
  hydrodynamic approach, these processes contribute weak decay terms
  to the continuity equations at local equilibrium, i.e., at the level
  of ``ideal'' hydrodynamics. Here we report the derivation of the
  decay term due to weak violation of energy conservation. Such terms
  have to be considered on equal footing with the well-known
  recombination terms due to nonconservation of the number of
  particles in each band. At high enough temperatures in the
  ``hydrodynamic regime'' supercollisions dominate both types of the
  decay terms (as compared to the leading-order electron-phonon
  interaction). We also discuss the contribution of supercollisions to
  the heat transfer equation (generalizing the continuity equation for
  the energy density in viscous hydrodynamics).
\end{abstract}

\maketitle

Electronic hydrodynamics is quickly growing into a mature field of
condensed matter physics \cite{pg,rev,luc}. Similarly to the usual
hydrodynamics \cite{dau6,chai}, this approach offers a universal,
long-wavelength description of collective flows in interacting
many-electron systems. As a macroscopic theory of strongly interacting
systems, hydrodynamics should appear to be extremely attractive for
condensed matter theorists dealing with problems where strong
correlations invalidate simple theoretical approaches. However,
electrons in solids exist in the environment created by a crystal
lattice and typically experience collisions with lattice imperfections
(or ``disorder'') and lattice vibrations (phonons). The former
typically dominate electronic transport at low temperatures, while at
high temperatures the electron-phonon interaction takes over. In both
cases the electron motion is diffusive (unless the sample size is
smaller than the mean free path in which case the motion is ballistic)
since in both types of scattering the electronic momentum is not
conserved. On the other hand, if a material would exist where the
momentum-conserving electron-electron interaction would dominate at
least in some non-negligible temperature range, then one could be
justified in neglecting the momentum non-conserving processes and
applying the hydrodynamic theory. In recent years, several extremely
pure materials became available with graphene being the most studied
\cite{pg,luc}.

As a manifestation of macroscopic conservation laws, hydrodynamics is
universal. Most conventional fluids are assumed Galilean invariant and
are described by the same set of hydrodynamic equations
\cite{dau6}. Similar approach was at the heart of the early
theoretical work on electronic hydrodynamics
\cite{gurzhi,Andreev2011}. Another well-known case is the relativistic
hydrodynamics \cite{dau6} relevant to neutron stars and interstellar
matter. Since its low-energy quasiparticles are characterized by the
Dirac spectrum, graphene attracted significant theoretical attention
as a possible condensed matter realization of relativistic
hydrodynamics \cite{har,mfs,mfss,mus,msf,job}. However, due to the
classical, three-dimensional nature of the Coulomb interaction between
electrons, the emergent hydrodynamics in graphene is neither
Galilean- nor Lorentz-invariant \cite{rev}.

In nearly neutral (or compensated) graphene the electron system is
non-degenerate (at least at relatively high temperatures where the
hydrodynamic approach is justified) with both the conductance and
valence bands contributing on equal footing. Although the electron
system is not Lorenz-invariant, the linearity of the Dirac spectrum
plays an important role. Firstly, the Auger processes are
kinematically suppressed leading to the near-conservation of the
number of particles in each band \cite{rev,luc,alf,bri}. Secondly, the
so-called collinear scattering singularity
\cite{msf,mfss,kash,mus,bri,schutt,hydro0,drag2} allows for a
non-perturbative solution to the kinetic (Boltzmann) equation focusing
on the three hydrodynamic modes \cite{hydro0,hydro1,me1}. As a result,
one can determine the general form of the hydrodynamic equations and
to evaluate the kinetic coefficients \cite{me1,me2,me3}. To be of any
practical value, the latter calculation has to be combined with the
renormalization group approach \cite{shsch} since the effective
coupling constant in real graphene (either encapsulated or put on a
dielectric substrate) is not too small, $\alpha_g\approx0.2-0.3$
\cite{sav,gal}.

Next to the conservation laws, the main assumption of the
hydrodynamic approach is local equilibrium
\cite{dau6,dau10} established by means of interparticle
collisions. Neglecting all dissipative processes, this allows (together
with the conservation laws) for a phenomenological derivation of
hydrodynamic equations \cite{dau6,chai} that can be further supported
by the kinetic theory, where the local equilibrium distribution
function nullifies the collision integral in the Boltzmann equation
\cite{dau10}. The resulting ideal hydrodynamics is described by the
Euler equation and the continuity equations. This is where the
electronic fluid in graphene differs from conventional fluids (both
Galilean- and Lorentz-invariant): as in any solid, conservation
laws in graphene are only approximate, leaving the collision integrals
describing scattering processes other than the electron-electron
interaction to be nonzero even in local equilibrium. This leads to the
appearance of weak decay terms in the continuity equations.

Two such terms have already been discussed in literature. Firstly,
even if the electron-electron interaction is the dominant scattering
process in the system, no solid is absolutely pure. Consequently, even
ultra-pure graphene samples possess some degree of weak
disorder. Disorder scattering violates momentum conservation and hence
a weak decay term must appear in the generalized Euler equation
\cite{rev,luc,hydro1,me1}. Secondly, conservation of the number of
particles in each band is violated by a number of processes (e.g., the
Auger and three-particle scattering). While commonly assumed to be
weak, they are manifested in the decay -- or recombination -- term in
the corresponding continuity equation. This was first established in
\cite{alf} in the context of thermoelectric phenomena (for the most
recent discussion see \cite{prin2020}). Later, quasiparticle
recombination was shown to lead to linear magnetoresistance in
compensated semimetals \cite{mr1,mrexp,mr3,mr2} as well as giant
magnetodrag \cite{drag12,meg}.

In this paper, we report the derivation of the third weak decay term
in the hydrodynamic theory in graphene due to weak violation of energy
conservation. Indeed, the electron-phonon interaction may lead not
only to the loss of electronic momentum (responsible for electrical
resistivity in most metals at high temperatures), but also to the loss
of energy. Although subdominant in the hydrodynamic regime, the
electron-phonon interaction should be taken into account as one of the
dissipative processes. In graphene, the linearity of the Dirac
spectrum once again plays an important role: as the speed of sound is
much smaller than the electron velocity $v_g$, leading-order
scattering on acoustic phonons is kinematically suppressed.
Consequently, scattering off the optical branch is usually considered
\cite{fos16,lev19}. In contrast, we argue that there is another
process, the disorder-assisted electron-phonon scattering \cite{srl}
or ``supercollisions'' \cite{ralph13,betz,tik18,kong}, that is
responsible for both quasiparticle recombination and energy
relaxation. In the high-temperature hydrodynamic regime, the
supercollisions are expected to dominate both decay contributions
\cite{srl}. Moreover, this process contributes weak decay terms to the
continuity equations already at local equilibrium, i.e., at the level
of ``ideal'' hydrodynamics.

Our arguments are based on the kinetic theory approach to electronic
transport. In the spirit of Ref.~\onlinecite{shsch}, we assume the
possibility of deriving the hydrodynamic equations from the kinetic
equation in the weak coupling limit \cite{me1}, $\alpha_g\ll1$, with
the subsequent renormalization of the kinetic coefficients to the
realistic parameter regime \cite{me2}. Under these assumptions, we
start with the kinetic equation
\begin{subequations}
\label{be}
\begin{equation}
\label{ke}
{\cal L}f_{\lambda\bs{k}} = {\rm St}_{ee} [f_{\lambda\bs{k}}] + {\rm St}_{R} [f_{\lambda\bs{k}}] 
+ {\rm St}_{\rm dis} [f_{\lambda\bs{k}}],
\end{equation}
with the Liouville's operator (in the left-hand side)
\begin{equation}
\label{l}
{\cal L} = \partial_t + \bs{v}\!\cdot\!\bs{\nabla}_{\bs{r}} + 
\left(e\bs{E} \!+\! \frac{e}{c} \bs{v}\!\times\!\bs{B}\right)\!\cdot\!\bs{\nabla}_{\bs{k}},
\end{equation}
\end{subequations}
and the collision integrals describing the electron-electron
interaction (${\rm St}_{ee}$), disorder scattering (${\rm St}_{\rm
  dis}$), and quasiparticle recombination (${\rm St}_{R}$), where in
this paper we focus on ``supercollisions''. We employ the following
notations for the Dirac spectrum (the chirality $\lambda=\pm 1$
distinguishes the conduction and valence bands)
\begin{subequations}
\label{eg}
\begin{equation}
\epsilon_{\lambda\bs{k}} = \lambda v_g k,
\end{equation}
and velocities
\begin{equation}
\label{vg}
\bs{v}_{\lambda\bs{k}}=\lambda v_g \frac{\bs{k}}{k}, \qquad
\bs{k} = \frac{\lambda k}{v_g}\bs{v}_{\lambda\bs{k}}
=\frac{\epsilon_{\lambda\bs{k}}\bs{v}_{\lambda\bs{k}}}{v_g^2}.
\end{equation}
\end{subequations}

Hydrodynamics is the macroscopic manifestation of the conservation of
energy, momentum, and the number of particles. In a two-band system,
the latter comprises excitations in both bands. In the conductance
band these are electron-like quasiparticles with the number density
($N=4$ reflects spin and valley degeneracy in graphene)
\begin{subequations}
\label{den}
\begin{equation}
\label{np}
n_+=N\!\int\!\!\frac{d^2k}{(2\pi)^2} f_{+,\bs{k}},
\end{equation} 
while in the valence band the quasiparticles are hole-like
\begin{equation}
\label{nm}
n_-=N\!\int\!\!\frac{d^2k}{(2\pi)^2} \left(1-f_{-,\bs{k}}\right),
\end{equation}
with the total ``charge'' (or ``carrier'') density being
\begin{equation}
\label{n}
n = n_+ - n_-.
\end{equation}
Assuming the numbers of particles in the conduction and valence bands
are conserved independently, we can also define the total
quasiparticle (``imbalance'' \cite{alf}) density
\begin{equation}
\label{ni}
n_I = n_+ + n_-.
\end{equation}
\end{subequations}
Global charge conservation (or gauge symmetry) can be expressed in
terms of the usual continuity equation. This can be obtained from
Eq.~(\ref{be}) by performing a summation over all quasiparticle
states upon which all three collision integrals vanish \cite{dau10}
\begin{subequations}
\label{gsym}
\begin{eqnarray}
&&
N\!\int\!\!\frac{d^2k}{(2\pi)^2} {\rm St}_{ee} [f_{\lambda\bs{k}}]
=
N\!\int\!\!\frac{d^2k}{(2\pi)^2} {\rm St}_{R} [f_{\lambda\bs{k}}] =
\\
&&
\nonumber\\
&&
\qquad\qquad\qquad\qquad
=
N\!\int\!\!\frac{d^2k}{(2\pi)^2} {\rm St}_{\rm dis} [f_{\lambda\bs{k}}]
=0.
\nonumber
\end{eqnarray}
As a result, the continuity equation has the usual form
\begin{equation}
\label{cen}
\partial_t n + \bs{\nabla}_{\bs{r}}\!\cdot\!\bs{j} = 0,
\end{equation}
\end{subequations}
where the corresponding current is defined as
\begin{equation}
\label{j}
\bs{j} \!=\! \bs{j}_+ \!-\! \bs{j}_- \!=\! 
N\!\!\int\!\!\frac{d^2k}{(2\pi)^2} 
\left[\bs{v}_{+,\bs{k}} f_{+,\bs{k}} \!-\! \bs{v}_{-,\bs{k}}\left(1\!-\!f_{-,\bs{k}}\right)\right],
\end{equation}

The rest of the conservation laws in graphene are approximate as
manifested in the collision integrals not vanishing upon corresponding
summations. The continuity equation expressing momentum conservation
(i.e. the Euler equation) is obtained by multiplying the kinetic
equation by the quasiparticle momentum $\bs{k}$ and summing over all
states. Since the electron-electron interaction conserves momentum, the 
corresponding collision integral vanishes
\begin{equation}
\label{mcon}
N\!\int\!\!\frac{d^2k}{(2\pi)^2} \bs{k} \, {\rm St}_{ee} [f_{\lambda\bs{k}}] =0.
\end{equation}
Weak disorder scattering is typically described within the simplest
$\tau$-approximation \cite{dau10}
\begin{equation}
\label{mdis}
N\!\int\!\!\frac{d^2k}{(2\pi)^2} \bs{k} \, {\rm St}_{\rm dis} [f_{\lambda\bs{k}}] 
= \frac{\bs{n}_{\bs{k}}}{\tau_{\rm dis}},
\end{equation}
where the momentum density is defined as
\begin{equation}
\label{nk}
\bs{n}_{\bs{k}} = N\!\sum_\lambda\!\int\!\frac{d^2k}{(2\pi)^2} \bs{k} f_{\lambda\bs{k}}
=
v_g^{-2}\bs{j}_E.
\end{equation}
The last equality represents the fact that in graphene the
momentum density is proportional to the energy density [due to the
properties of the Dirac spectrum Eq.~(\ref{eg})].

Supercollisions contributing to the recombination collision integral
also violate momentum conservation, however, in comparison to the above
weak disorder scattering, this is a second-order process. Moreover,
the disorder mean free time $\tau_{\rm dis}$ is typically determined
from experimental data (see e.g., Ref.~\onlinecite{gal}) and hence can
be assumed to include the contribution of supercollisions as well.

The remaining two continuity equations -- energy and quasiparticle
imbalance -- are unaffected by the electron-electron interaction and
weak disorder scattering. Indeed, the electron-electron interaction
conserves energy and -- neglecting the Auger processes -- particle
number in each band:
\begin{equation}
\label{ciee}
N\!\int\!\!\frac{d^2k}{(2\pi)^2} \epsilon_{\lambda\bs{k}} {\rm St}_{ee} [f_{\lambda\bs{k}}] 
=
N\!\int\!\!\frac{d^2k}{(2\pi)^2} \lambda \, {\rm St}_{ee} [f_{\lambda\bs{k}}]
=0.
\end{equation}
Same applies to the (elastic) disorder scattering
\begin{equation}
\label{cidis}
N\!\int\!\!\frac{d^2k}{(2\pi)^2} \epsilon_{\lambda\bs{k}} {\rm St}_{\rm dis} [f_{\lambda\bs{k}}] 
=
N\!\int\!\!\frac{d^2k}{(2\pi)^2} \lambda \, {\rm St}_{\rm dis} [f_{\lambda\bs{k}}]
=0.
\end{equation}
However, supercollisions do not conserve both quantities and hence
lead to weak decay terms in the two continuity equations.

Let us now specify the collision integral describing the
disorder-assisted electron-phonon scattering. An electron in the upper
(conductance) band may scatter into an empty state in the lower
(valence) band -- effectively recombining with a hole -- emitting a
phonon (which carries away the energy) and losing its momentum to the
impurity. Within the standard approach to the electron-phonon
interaction, this process is described by the collision integral
\begin{subequations}
\label{ceph1}
\begin{eqnarray}
\label{ir+1}
&&
{\rm St}_{R} [f_{+\bs{k}_2}] = 2\pi\sum_{\bs{k}_1,\bs{q}} 
W_q 
\delta(\epsilon_{+\bs{k}_2}\!-\epsilon_{-\bs{k}_1}\!-\omega_q)
\times
\\
&&
\nonumber\\
&&
\qquad
\times
\left[ f_{-\bs{k}_1}(1\!-\!f_{+\bs{k}_2})n_q
-
f_{+\bs{k}_2} (1\!-\!f_{-\bs{k}_1})(1\!+\!n_q)\right],
\nonumber
\end{eqnarray}
where $n_q$ is the phonon (Plank's) distribution function (the phonons
are assumed to be at equilibrium and play the role of a ``bath''),
$W_q$ is the effective scattering rate that includes the Dirac factors
and is averaged over the angles \cite{srl}.

Similarly, an electron in the lower band may absorb a phonon and
scatter into the upper band -- effectively creating an electron-hole
pair -- while still losing its momentum to the impurity
\begin{eqnarray}
\label{ir-1}
&&
{\rm St}_{R} [f_{-\bs{k}_2}] = 2\pi\sum_{\bs{k}_1,\bs{q}} 
W_q 
\delta(\epsilon_{+\bs{k}_1}\!-\epsilon_{-\bs{k}_2}\!-\omega_q)
\times
\\
&&
\nonumber\\
&&
\qquad
\times
\left[ f_{+\bs{k}_1}(1\!-\!f_{-\bs{k}_2})(1\!+\!n_q)
-
f_{-\bs{k}_2} (1\!-\!f_{+\bs{k}_1})n_q\right].
\nonumber
\end{eqnarray}

The collision integral (\ref{ceph1}) conserves the total charge
\begin{equation}
\label{ncon}
N\sum_{\bs{k}} {\rm St}_{R} [f_{\lambda\bs{k}}] 
=
N\sum_{\bs{k}_2}\left({\rm St}_{R} [f_{+\bs{k}_2}] \!+\! {\rm St}_{R} [f_{-\bs{k}_2}]\right)
=0,
\end{equation}
[see Eq.~(\ref{gsym})] and vanishes in global equilibrium
\begin{equation}
\label{eq0}
{\rm St}_{R}[f^{(0)}]=0,
\end{equation}
\end{subequations}
where the quasiparticle distribution is described by the Fermi
function. This should be contrasted with {\it local equilibrium}
described by
\begin{equation}
\label{le}
f^{(le)}_{\lambda\bs{k}} =
\left\{
1\!+\!\exp\left[\frac{\epsilon_{\lambda\bs{k}}\!-\!\mu_\lambda(\bs{r}) \!-\! 
\bs{u}(\bs{r})\!\cdot\!\bs{k}}{T(\bs{r})}\right]\!
\right\}^{\!-1}\!,
\end{equation}
where ${\mu_\lambda(\bs{r})}$ is the local chemical potential and
$\bs{u}(\bs{r})$ is the hydrodynamic (or ``drift'') velocity. The
local equilibrium distribution function (\ref{le}) allows for
independent chemical potentials in the two bands, which can be
expressed in terms of the ``thermodynamic'' and ``imbalance'' chemical
potentials
\begin{equation}
\label{mui}
\mu_\lambda = \mu + \lambda\mu_I.
\end{equation}
In global equilibrium
\begin{equation}
\label{ff}
f^{(0)} = f^{(le)}_{\lambda\bs{k}}(\mu_I=0, \bs{u}=0).
\end{equation}
Now we show, that in local equilibrium, i.e. for nonzero $\mu_I$, the
recombination collision integral remains finite [unlike
  Eq.~(\ref{eq0})]. As a scalar quantity, the collision integral
(\ref{ceph1}) cannot depend on the hydrodynamic velocity $\bs{u}$ in
the first (linear) order. Consequently, to the leading order the
integrated collision integral yielding the decay terms in the
continuity equations is proportional to $\mu_I$.

To the leading order, we can describe the difference between the local
equilibrium distribution function $f^{(le)}_{\lambda\bs{k}}$ and the
Fermi function $f^{(0)}$ similarly to the leading non-equilibrium
correction in the standard derivation of hydrodynamics \cite{dau10}
\begin{equation}
\label{df}
\delta f = 
f_{\lambda\bs{k}} \!-\! f^{(0)}
=
- T \frac{\partial f^{(0)}}{\partial\epsilon} h
=
f^{(0)} \left(1\!-\!f^{(0)}\right) h.
\end{equation}
Now we re-write the collision integral (\ref{ceph1}) with the help of
the relations
\begin{eqnarray*}
&&
f_1(1\!-\!f_{2})(1\!+\!n_q)-f_{2} (1\!-\!f_{1})n_q
=
\\
&&
\\
&&
\qquad
=
(1\!-\!f_{1})(1\!-\!f_{2})(1\!+\!n_q)
\left[\frac{f_1}{1\!-\!f_1} \frac{n_q}{1\!+\!n_q} - \frac{f_2}{1\!-\!f_2}\right],
\end{eqnarray*}
and
\[
f_{2}^{(0)}\left(1\!-\!f_1^{(0)}\right)(1\!+\!n_q)
=
-\frac{\partial n_q}{\partial\omega}
\left(f_{2}^{(0)}\!-\!f_1^{(0)}\right),
\]
and find (to the leading order in $h$)
\begin{subequations}
\label{ceph2}
\begin{eqnarray}
\label{ir+2}
&&
{\rm St}_{R} [f_{+\bs{k}_2}] = -2\pi\sum_{\bs{k}_1,\bs{q}} 
W_q 
\delta(\epsilon_{+\bs{k}_2}\!-\epsilon_{-\bs{k}_1}\!-\omega_q)
\frac{\partial n_q}{\partial\omega}
\nonumber\\
&&
\nonumber\\
&&
\qquad\qquad
\times
\left(f_{+\bs{k}_2}^{(0)}\!-\!f_{-\bs{k}_1}^{(0)}\right)
(h_{-\bs{k}_1}\!-\!h_{+\bs{k}_2}),
\end{eqnarray}
\begin{eqnarray}
\label{ir-2}
&&
{\rm St}_{R} [f_{-\bs{k}_2}] = 2\pi\sum_{\bs{k}_1,\bs{q}} 
W_q 
\delta(\epsilon_{-\bs{k}_2}\!-\epsilon_{+\bs{k}_1}\!+\omega_q)
\frac{\partial n_q}{\partial\omega}
\nonumber\\
&&
\nonumber\\
&&
\qquad\qquad
\times
\left(f_{+\bs{k}_1}^{(0)}\!-\!f_{-\bs{k}_2}^{(0)}\right)
(h_{-\bs{k}_2}\!-\!h_{+\bs{k}_1}).
\end{eqnarray}
\end{subequations}

Consider now the contribution of the recombination collision integral
to the continuity equation for the quasiparticle imbalance
\begin{eqnarray}
\label{cephi}
&&
N\sum_{\bs{k}}\lambda \, {\rm St}_{R} [f_{\lambda\bs{k}}] 
=
N\sum_{\bs{k}_2}\left({\rm St}_{R} [f_{+\bs{k}_2}] \!-\! {\rm St}_{R} [f_{-\bs{k}_2}]\right)
=
\nonumber\\
&&
\nonumber\\
&&
\qquad
=
-4\pi N\!\sum_{\bs{k}_1,\bs{k}_2,\bs{q}} \!
W_q 
\delta(\epsilon_{+\bs{k}_2}\!-\epsilon_{-\bs{k}_1}\!-\omega_q)
\frac{\partial n_q}{\partial\omega}
\nonumber\\
&&
\nonumber\\
&&
\qquad\qquad
\times
\left(f_{+\bs{k}_2}^{(0)}\!-\!f_{-\bs{k}_1}^{(0)}\right)
(h_{-\bs{k}_1}\!-\!h_{+\bs{k}_2}).
\end{eqnarray}
To the leading order, the deviation $h_{\lambda\bs{k}}$ is
proportional to $\mu_I$
\begin{equation}
\label{hle}
h_{\lambda\bs{k}} \approx \frac{\lambda\mu_I}{T}.
\end{equation}
The remaining integral has dimensions of particle density divided by
time and therefore the result can be written in two equivalent forms
\begin{eqnarray}
\label{cephir}
N\sum_{\bs{k}}\lambda \, {\rm St}_{R} [f_{\lambda\bs{k}}] 
\approx
-
\mu_I n_{I,0} \lambda_Q
\approx
-
\frac{n_I\!-\!n_{I,0}}{\tau_R}.
\end{eqnarray}
Here $n_I$ is the imbalance density (\ref{ni}) in local equilibrium,
while $n_{I,0}$ is the same quantity evaluated with the Fermi
distribution function (\ref{ff}), i.e. for $\mu_I=0$ and
$\bs{u}=0$. The first equality in Eq.~(\ref{cephir}) coincides with
the expression used in Ref.~\onlinecite{alf} and serves as the
definition of the dimensionless coefficient $\lambda_Q$, while the
second (valid to the leading order) was suggested in
Refs.~\onlinecite{me1,mr1} and provides the definition of the
``recombination time'' $\tau_R$ (see also
Ref.~\onlinecite{prin2020}). The two expressions are equivalent since
${n_I\!-\!n_{I,0}\propto\mu_I}$.

The same scattering process contributes a weak decay term to the
continuity equation for the energy density. Indeed, multiplying the
collision integral (\ref{ceph1}) by the quasiparticle energy and
summing over all states, we find after similar algebra
\begin{eqnarray}
\label{cephe}
&&
N\sum_{\bs{k}}\epsilon_{\lambda\bs{k}} {\rm St}_{R} [f_{\lambda\bs{k}}] 
=
\\
&&
\nonumber\\
&&
\quad
=
N\sum_{\bs{k}_2}\left(\epsilon_{+\bs{k}_2}{\rm St}_{R} [f_{+\bs{k}_2}] 
\!+\!\epsilon_{-\bs{k}_2} {\rm St}_{R} [f_{-\bs{k}_2}]\right)
=
\nonumber\\
&&
\nonumber\\
&&
\quad
=
-2\pi N\!\sum_{\bs{k}_1,\bs{k}_2,\bs{q}} 
W_q 
\delta(\epsilon_{+\bs{k}_2}\!-\epsilon_{-\bs{k}_1}\!-\omega_q)
\frac{\partial n_q^{(0)}}{\partial\omega}\omega_q
\nonumber\\
&&
\nonumber\\
&&
\qquad\qquad
\times
\left(f_{+\bs{k}_2}^{(0)}\!-\!f_{-\bs{k}_1}^{(0)}\right)
(h_{-\bs{k}_1}\!-\!h_{+\bs{k}_2}).
\nonumber
\end{eqnarray}
Defining the decay coefficients similarly to Eq.~(\ref{cephir}) above,
we may present the result in the form
\begin{equation}
\label{cepher}
N\sum_{\bs{k}}\epsilon_{\lambda\bs{k}} {\rm St}_{R} [f_{\lambda\bs{k}}]
=
-
\mu_I n_{E,0} \lambda_{QE}
\approx
-
\frac{n_E\!-\!n_{E,0}}{\tau_{RE}}.
\end{equation}
Here the equivalence of the two forms of the decay term stems from the
fact that ${n_E\!-\!n_{E,0}\propto\mu_I}$ assuming the electrons and
holes are characterized by the same temperature.

Supercollisions are not the only scattering process contributing to
both quasiparticle recombination and energy relaxation. Clearly,
direct (not impurity-assisted) electron-phonon interaction contributes
to energy relaxation as well as to quasiparticle recombination (in the
case of intervalley scattering) \cite{meg,hydro0,srl,alf,drag2}. In
addition, optical phonons may also contribute \cite{fos16,lev19},
although within the hydrodynamic approach these contributions were
discussed only at the level of dissipative (viscous) hydrodynamics
\cite{lev19}. The contribution of the direct \cite{bistr,tse} and
impurity assisted electron-phonon scattering to energy relaxation was
compared in \cite{srl}, where it was argued that at high enough
temperatures, $T\gtrsim T_{BG}$ (where $T_{BG}$ is the
Bloch-Gr\"uneisen temperature) the supercollisions do dominate. In the
degenerate regime, where $T_{BG}\propto\sqrt{n}$, Ref.~\cite{srl}
estimates $T_{BG}$ as ``few tens of Kelvin''. At charge neutrality, we
estimate $T_{BG}=(s/v_g)T\ll T$ (where $s$ is the speed of sound),
such that supercollisions always dominate over direct electron-phonon
coupling. Moreover, taking into account the additional scattering
processes will not change the form of the decay terms (\ref{cephir})
and (\ref{cepher}), but rather change the numerical values of the
parameters $\lambda_Q$ and $\lambda_{QE}$, which may have to be
considered phenomenological while interpreting experimental data
\cite{meg}.

At charge neutrality and in the hydrodynamic regime, the coefficients
$\lambda_Q$ and $\lambda_{QE}$ are of the same order of magnitude
(both are dominated by the same supercollisions), but quantitatively
different. Indeed, the continuity equation for the energy density
should contain one more contribution of the similar form.
``Quasiparticle recombination'' typically refers to scattering between
the quasiparticle states in different bands only. This is the only
type of supercollisions contributing to $\lambda_Q$. Similar
supercollisions may also take place within a single band \cite{srl}.
While this process does not change the number of particles in the
band, it does describe the energy loss as the electron may scatter
from the higher energy state into the lower energy state (and losing
its momentum to the impurity along the way). Consequently, this
additional process does contribute to energy relaxation. Given that
the form of the corresponding collision integral is very similar to
Eq.~(\ref{ceph1}) -- one only has to change to band indices to be the
same -- the algebra remains the same and thus we can treat
Eq.~(\ref{cepher}) as the final result that takes this additional
intraband supercollisions into account making the numerical values of
$\lambda_Q$ and $\lambda_{QE}$ substantially different -- we do not
expect any accidental cancellation or smallness should the difference
$\lambda_Q-\lambda_{QE}$ appear in a particular solution of
hydrodynamic equations. At the same time, at low temperatures -- i.e.,
below the hydrodynamic range -- we expect the coefficients $\lambda_Q$
and $\lambda_{QE}$ to be parametrically different: energy relaxation
is now dominated by the direct electron-phonon interaction \cite{srl},
while the recombination is still governed by supercollisions (together
with the phonon-induced intervalley scattering).

The order of magnitude of $\tau_R$ could be estimated based on the
calculations of Ref.~\cite{srl}. Adapting the latter to charge
neutrality, we find $\tau_R^{-1}\sim D^2T^2/(\rho s^2 v_g^2)$ (where
$D\approx 20\,$eV is the deformation potential \cite{bistr,tse} and
$\rho$ is the mass density per unit area) yielding the corresponding
length scale $\ell_R^{-1}\approx 10 \,\mu$m at the top of the
hydrodynamic temperature range, $T\approx250\,$K. This should be
further compared to the contribution of three-particle collisions
\cite{luc,lev19}, $\tau_3^{-1}\sim\alpha_g^4T$. Assuming the common
sample design where graphene is encapsulated in hexagonal boron
nitride (with the dielectric constant $\epsilon\approx4$), the
effective coupling constant (taking into account renormalizations) is
$\alpha_g\approx0.3-0.4$ leading to the similar estimate at high
temperatures. On the other hand, at the low end of the hydrodynamic
range \cite{geim3}, $T\approx50\,$K, the contribution of the
three-body collisions should dominate (accounting for the empirical
value $\ell_R\approx1.2\,\mu$m reported in \cite{meg}), however
preserving the functional form of the weak decay terms in the
continuity equations.

Finally, once the dissipative processes due to electron-electron
interaction are taken into account, one usually replaces the
continuity equation for the energy density by the equivalent equation
for the entropy density, the so-called ``heat transfer equation''
\cite{dau6}. The decay terms discussed in this paper appear in that
equation as well. Let us briefly discuss their form.

Recall the derivation of the continuity equation for the entropy from
the kinetic equation \cite{me1}. The entropy density of a system of fermions is
defined in terms of the distribution function as
\begin{equation}
\label{sdk}
s \!=\! -\! N\!\sum_\lambda\!\!\int\!\!\frac{d^2k}{(2\pi)^2} 
\!\left[f_{\lambda\bs{k}} \ln f_{\lambda\bs{k}}
\!+\! (1\!-\!f_{\lambda\bs{k}})\!\ln(1\!-\!f_{\lambda\bs{k}})\right]\!.
\end{equation}
Treating this integral as 
\[
s = N\!\sum_\lambda\!\int\!\frac{d^2k}{(2\pi)^2} {\cal S}[f_{\lambda\bs{k}}],
\]
any derivative of $s$ can be represented in the form
\[
\frac{\partial s}{\partial z} = 
N\!\sum_\lambda\!\int\!\frac{d^2k}{(2\pi)^2} 
\frac{\partial {\cal S}[f_{\lambda\bs{k}}]}{\partial f_{\lambda\bs{k}}} 
\frac{\partial f_{\lambda\bs{k}}}{\partial z}.
\]

Consider now each term of the kinetic equation multiplied by the
derivative $\partial {\cal S}[f_{\lambda\bs{k}}]/\partial
f_{\lambda\bs{k}}$ and summed over all states. Using the above
relation with ${z\rightarrow{t}}$, one finds for the time derivative
term
\[
N\!\sum_\lambda\!\int\!\frac{d^2k}{(2\pi)^2} 
\frac{\partial {\cal S}[f_{\lambda\bs{k}}]}{\partial f_{\lambda\bs{k}}} 
\frac{\partial f_{\lambda\bs{k}}}{\partial t}
=
\frac{\partial s}{\partial t}.
\]
The gradient term yields similarly
\begin{eqnarray*}
&&
N\!\sum_\lambda\!\int\!\frac{d^2k}{(2\pi)^2} 
\frac{\partial {\cal S}[f_{\lambda\bs{k}}]}{\partial f_{\lambda\bs{k}}}
\bs{v}_{\lambda\bs{k}}\cdot\bs{\nabla}_{\bs{r}}f_{\lambda\bs{k}}
=
\\
&&
\\
&&
\qquad\qquad
=
\bs{\nabla}_{\bs{r}}\cdot
N\!\sum_\lambda\!\int\!\frac{d^2k}{(2\pi)^2}
\bs{v}_{\lambda\bs{k}}{\cal S}[f_{\lambda\bs{k}}]
=
\bs{\nabla}_{\bs{r}}\!\cdot\!\bs{j}_S,
\end{eqnarray*}
where the quantity
\begin{equation}
\label{jen0}
\bs{j}_S =  N\!\sum_\lambda\!\int\!\frac{d^2k}{(2\pi)^2} \bs{v}_{\lambda\bs{k}}
{\cal S}[f_{\lambda\bs{k}}],
\end{equation}
can be interpreted as the entropy current.

The electric field term vanishes as the total derivative
\begin{eqnarray*}
&&
e\bs{E}\cdot
N\!\sum_\lambda\!\int\!\frac{d^2k}{(2\pi)^2} 
\frac{\partial {\cal S}[f_{\lambda\bs{k}}]}{\partial f_{\lambda\bs{k}}}
\bs{\nabla}_{\bs{k}}f_{\lambda\bs{k}}
=
\\
&&
\\
&&
\qquad\qquad
=
e\bs{E}\cdot
N\!\sum_\lambda\!\int\!\frac{d^2k}{(2\pi)^2} 
\bs{\nabla}_{\bs{k}}{\cal S}[f_{\lambda\bs{k}}]
=0,
\end{eqnarray*}
while the Lorentz term vanishes for rotationally invariant systems
upon integrating by parts [justified by the fact that
${{\cal{S}}(k\rightarrow\infty)\rightarrow0}$]
\begin{eqnarray*}
&&
\frac{e}{c}
N\!\sum_\lambda\!\int\!\frac{d^2k}{(2\pi)^2} 
\frac{\partial {\cal S}[f_{\lambda\bs{k}}]}{\partial f_{\lambda\bs{k}}}
\left[\bs{v}_{\lambda\bs{k}}\!\times\!\bs{B}\right]\!\cdot\!\bs{\nabla}_{\bs{k}}f_{\lambda\bs{k}}
=
\\
&&
\\
&&
\qquad\qquad
=
\frac{e}{c}
N\!\sum_\lambda\!\int\!\frac{d^2k}{(2\pi)^2} 
\left[\bs{v}_{\lambda\bs{k}}\!\times\!\bs{B}\right]\!\cdot\!\bs{\nabla}_{\bs{k}}{\cal S}[f_{\lambda\bs{k}}]
\\
&&
\\
&&
\qquad\qquad
=
-\frac{e}{c}
N\!\sum_\lambda\!\int\!\frac{d^2k}{(2\pi)^2} 
{\cal S}[f_{\lambda\bs{k}}]\bs{\nabla}_{\bs{k}}\!\cdot\!\left[\bs{v}_{\lambda\bs{k}}\!\times\!\bs{B}\right]
=0.
\end{eqnarray*}
The last equality follows from 
\[
\frac{\partial v^\alpha_{\lambda\bs{k}}}{\partial k^\beta}
=\frac{v_g}{\lambda k}\left(\delta_{\alpha\beta}-\frac{k^\alpha k^\beta}{k^2}\right).
\]
Similar approach was used in Ref.~\onlinecite{me1} to derive
the continuity equations (as outlined above).

Combining all four terms, we conclude that integration with
the factor
${\partial{\cal{S}}[f_{\lambda\bs{k}}]/\partial{f}_{\lambda\bs{k}}}$
turns the left-hand side of the kinetic equation to the familiar form
\begin{equation}
\label{lhsent}
N\!\sum_\lambda\!\int\!\frac{d^2k}{(2\pi)^2} 
\frac{\partial {\cal S}[f_{\lambda\bs{k}}]}{\partial f_{\lambda\bs{k}}}
{\cal L}f_{\lambda\bs{k}}
=
\frac{\partial s}{\partial t} + \bs{\nabla}_{\bs{r}}\!\cdot\!\bs{j}_S.
\end{equation}
Eq.~(\ref{lhsent}) is valid for an arbitrary distribution function.

Denoting the integral of the right-hand side of the kinetic equation
by
\begin{equation}
\label{rhsent}
{\cal I} = N\!\sum_\lambda\!\!\int\!\!\frac{d^2k}{(2\pi)^2} 
\frac{\partial {\cal S}[f_{\lambda\bs{k}}]}{\partial f_{\lambda\bs{k}}} 
\!\left({\rm St}_{ee} [f] + {\rm St}_{R} [f] + {\rm St}_{\rm dis} [f] 
\right)\!,
\end{equation}
we arrive at the ``continuity equation for the entropy''
\begin{equation}
\label{ceen}
\frac{\partial s}{\partial t} + \bs{\nabla}_{\bs{r}}\!\cdot\!\bs{j}_S 
=
{\cal I}.
\end{equation}
In the usual hydrodynamics \cite{dau6} the only contribution to the
collision integral is given by particle-particle scattering, i.e. the
processes assumed to be responsible for establishing local equilibrium
such that at ${{\cal I}=0}$ the ideal (Euler) hydrodynamic is
isentropic. In the present case, local equilibrium is assumed to be
achieved by means of the electron-electron interaction.

Evaluating the derivative of ${\cal S}$ explicitly, we find
\[
\frac{\partial {\cal S}[f_{\lambda\bs{k}}]}{\partial f_{\lambda\bs{k}}}
=
- \ln \frac{f_{\lambda\bs{k}}}{1\!-\!f_{\lambda\bs{k}}}
=
\ln \left[\frac{1}{f_{\lambda\bs{k}}}\!-\!1\right].
\]
For the local equilibrium distribution function
\[
\frac{\partial {\cal S}[f_{\lambda\bs{k}}]}{\partial f_{\lambda\bs{k}}}
=
\frac{\epsilon_{\lambda\bs{k}}\!-\!\mu_\lambda \!-\! 
\bs{u}\!\cdot\!\bs{k}}{T}.
\]
Substituting this expression into Eq.~(\ref{rhsent}), we find that the
remaining integration is very similar to the above derivation of the
continuity equations.

The integral with the quasiparticle energy yields exactly the above
Eq.~(\ref{cepher}). The integral with $\lambda\mu_I$ yields
Eq.~(\ref{cephir}) multiplied by $\mu_I$. Finally, the term
$\bs{u}\!\cdot\!\bs{k}$ yields Eq.~(\ref{mdis}) multiplied by the
hydrodynamic velocity. The integral of this term with the
recombination collision integral is assumed to be included into the
definition of the mean free time, see the corresponding discussion
above. As a result, we arrive at the following form of the integrated
collision integral
\begin{eqnarray}
\label{rhsent2}
&&
{\cal I}=
-\frac{1}{T} \frac{n_E\!-\!n_{E,0}}{\tau_{RE}}+
\frac{\mu_I}{T} \frac{n_I\!-\!n_{I,0}}{\tau_R}+
\frac{\bs{u}\!\cdot\!\bs{n}_{\bs{k}}}{T\tau_{\rm dis}}.
\nonumber
\end{eqnarray}

The decay terms (\ref{rhsent2}) appear already at local
equilibrium. To complete the heat transfer equation one has to take
into account dissipation. In graphene, this is most conveniently done
by considering the classical limit of relativistic hydrodynamics since
the form of dissipative corrections is determined by the symmetries of
the quasiparticle spectrum. The result has been already reported in
literature, therefore we combine the dissipative corrections with
Eq.~(\ref{rhsent2}) and write the heat transfer equation in graphene
in the form (here $\delta\bs{j}$ and $\delta\bs{j}_I$ are the
dissipative corrections to the electric and imbalance currents,
respectively).
\begin{eqnarray}
\label{eqent}
&&
T\left[\frac{\partial s}{\partial t}
+
\bs{\nabla}_{\bs{r}}\!\cdot\!
\left(s\bs{u}-\delta\bs{j}\frac{\mu}{T}-\delta\bs{j}_I\frac{\mu_I}{T}\right)\right]
=
\\
&&
\nonumber\\
&&
\qquad\qquad
=
\delta\bs{j}\!\cdot\!
\left[e\bs{E}\!+\!\frac{e}{c}\bs{u}\!\times\!\bs{B}\!-\!T\bs{\nabla}\frac{\mu}{T}\right]
-
T\delta\bs{j}_I\!\cdot\!\bs{\nabla}\frac{\mu_I}{T}
\nonumber\\
&&
\nonumber\\
&&
\qquad\qquad\qquad
+
\frac{\eta}{2}\left(\nabla_\alpha u_\beta \!+\! \nabla_\beta u_\alpha
\!-\! \delta_{\alpha\beta} \bs{\nabla}\!\cdot\!\bs{u}\right)^2
\nonumber\\
&&
\nonumber\\
&&
\qquad\qquad\qquad
-
\frac{n_E\!-\!n_{E,0}}{\tau_{RE}}
+
\mu_I \frac{n_I\!-\!n_{I,0}}{\tau_R}
+
\frac{\bs{u}\!\cdot\!\bs{n}_{\bs{k}}}{\tau_{\rm dis}}.
\nonumber
\end{eqnarray}
The obtained equation (\ref{eqent}) should be compared to the
corresponding equations in Refs.~\onlinecite{luc,alf,lev19}, where
energy relaxation due to supercollisions were not taken into
account. All other terms are present in all four equations with the
following exceptions. The equation (54) of Ref.~\onlinecite{luc} is
written in the relativistic notation omitting the imbalance mode,
quasiparticle recombination, and disorder scattering, all of which are
discussed separately elsewhere in Ref.~\onlinecite{luc}.
Ref.~\onlinecite{alf} was the first to focus on the imbalance mode
with the equation (2.6) containing all the terms of Eq.~(\ref{eqent})
except for the viscous term (and energy relaxation). Finally, the
equation (1c) of Ref.~\onlinecite{lev19} contains all of the terms in
Eq.~(\ref{eqent}) except for energy relaxation and in addition
contains a term describing energy relaxation due to electrons
scattering on the optical phonon branch that is neglected here
(generalization of the resulting theory is straightforward).

To summarize, we have considered supercollisions as a mechanism of
quasiparticle recombination and energy relaxation in graphene and
derived the corresponding decay terms in the hydrodynamic continuity
equations. Since the same scattering mechanism is responsible for both
effects, one has to take into account energy relaxation while
considering quasiparticle recombination. The latter is an
indispensable feature of electronic hydrodynamics in graphene in
constrained geometries, where homogeneous solutions violate the
boundary conditions \cite{mr1}.

\section*{Acknowledgments} 

The authors are grateful to P. Alekseev, A. Andreev, U. Briskot,
A. Dmitriev, L. Golub, V. Kachorovskii, E. Kiselev, A. Mirlin,
J. Schmalian, M. Sch\"utt, A. Shnirman, K. Tikhonov, and M. Titov for
fruitful discussions. This work was supported by the German Research
Foundation DFG within FLAG-ERA Joint Transnational Call (Project
GRANSPORT), by the European Commission under the EU Horizon 2020
MSCA-RISE-2019 program (Project 873028 HYDROTRONICS), by the
German-Israeli Foundation for Scientific Research and Development
(GIF) Grant no. I-1505-303.10/2019 (IVG) and by the Russian Science
Foundation, Grant No. 20-12-00147 (IVG). BNN acknowledges the support
by the MEPhI Academic Excellence Project, Contract No. 02.a03.21.0005.


\bibliography{viscosity_refs}

\end{document}